\documentstyle[12pt,epsfig]{article}
  \def\lsim{\raise0.3ex\hbox{$<$\kern-0.75em\raise-1.1ex\hbox{$\sim$}}}
\def\gsim{\raise0.3ex\hbox{$>$\kern-0.75em\raise-1.1ex\hbox{$\sim$}}}
\def\noi{\noindent}

\def\bea{\begin{eqnarray}}  \def\eea{\end{eqnarray}}
\def\beq{\begin{equation}}   \def\eeq{\end{equation}}

\def\beeq{\begin{eqnarray}} \def\eeeq{\end{eqnarray}}

\begin{document}
\vbox to 1 truecm {}
\begin{center}
{\large \bf BARYON AND ANTIBARYON PRODUCTION IN HADRONIC AND NUCLEAR INTERACTIONS} \\ 
\vskip 8 truemm
{\bf A. Capella, E. G. Ferreiro and C. A. Salgado} \\
Laboratoire de Physique Th\'eorique et Hautes Energies\footnote{Laboratoire associ\'e
au Centre National de la Recherche Scientifique - URA D0063} \\
Universit\'e de Paris XI, B\^atiment 210,
F-91405 Orsay Cedex, France \\
\end{center}

\vskip 1 truecm

\begin{abstract}
We introduce a new formulation of the diquark breaking mechanism which describes with no free
parameters the huge nuclear stopping observed in central nucleus-nucleus collisions.
Supplemented, in the dual parton model, with strings originating from diquark--antidiquark pairs
from the nucleon sea, it gives a substantial increase of hyperon and antihyperon yields from $pPb$
to central $PbPb$ collisions. Compared to data, this increase is slightly underestimated for
$\Lambda$'s and $\Xi$'s but is five times too small for $\Omega$'s. Introducing final state
interactions, a reasonable description of all baryon and antibaryon yields is achieved. 
\end{abstract}

\vskip 2 truecm

\noi LPTHE Orsay 99/06 \par
\noi January 1999

\newpage
\pagestyle{plain}
\baselineskip=24 pt
Enhancement of strange baryon and anti-baryon production in central nucleus-nucleus collisions has
been observed experimentally \cite{1r} \cite{2r}. In particular, a spectacular enhancement of $\Xi$'s
and $\Omega$'s from $pPb$ to central $PbPb$ collisions has been measured \cite{2r}. This phenomenon
has been proposed as a signal of quark gluon plasma formation \cite{3r}. On the other hand
explanations based on string models have also been proposed \cite{4r}. In ref. \cite{5r} enhancement
of $\Lambda$'s and $\bar{\Lambda}$'s was described in the dual parton model (DPM) \cite{6r} with 
final state interaction. An important drawback of \cite{6r} is the lack of nucleon stopping --which
would increase the net baryon yield at mid rapidities. As a consequence a very strong effect due to
final state interaction was needed, as well as a rather large value of the fraction of strange
quarks in the nucleon sea. A mechanism of nuclear stopping based on diquark breaking (DB) has been
known for a long time \cite{7r}-\cite{9r}. Recently, it has been introduced in nucleus-nucleus
interactions \cite{10r}-\cite{11r}.\par

The purpose of this paper is threefold. First, we propose a new formulation of the DB stopping
mechanism of \cite{10r}-\cite{11r}, which allows to describe net baryon stopping without any
dynamical parameter governing the DB probability. Second, we show that the 
DPM with DB and
strings with diquarks--antidiquark pairs ($d$-$\bar{d}$) from the sea at their ends, allows to des\-cri\-be most of the
observed yields of $\Lambda$, $\bar{\Lambda}$, $\Xi$ and $\bar{\Xi}$ --but
significantly underestimates the production of $\Omega$'s and $\bar{\Omega}$'s. Third, using the
cor\-res\-pon\-ding baryon densities as initial conditions for final state interaction we achieve a
reasonable description of all baryon and antibaryon production\footnote{In this paper we
concentrate ourselves on the rapidity density. For a calculation of the transverse mass spectra in
DPM see \protect{\cite{12r}}.}. \par \vskip 5 truemm

\noi {\bf \underbar{Baryon stopping}.} We describe baryon stopping via the mechanism shown in
Fig.~1 --the same as Fig.~5 of \cite{11r}. However, contrary to \cite{11r} we assume here that there
is no dynamical suppression associated to DB --i.e. the weight of the DB diagram of Fig.~1 is the
same as the conventional DP one of Fig.~2. More precisely, in the case of a single inelastic
collision per nucleon (two string configuration), two valence quarks of the nucleon are sitting
together at a string end. We assume that they form a sort of compact object which hadronizes in the
conventional way (DP mechanism). However, in the presence of several inelastic
collisions we assume that the net baryon number (string junction) has equal probability of finding
itself in any of them. If it follows the valence diquark (Fig.~2) we have the conventional DP
situation. In the other cases it joins with sea quarks producing a much slower baryon (DB mechanism
of Fig.~1). The rapidity distribution of the net
baryon production $\Delta B = B - \bar{B}$ in $AA$ collisions is then given by    

\beq
\label{1e}
{dN^{AA\to \Delta B} \over dy} (y) = {\bar{n}_A \over \bar{n}} \left [ \bar{n}_A \left (
{dN^{\Delta B}_{DP} \over dy}(y) \right )_{\bar{n}/\bar{n}_A} + (\bar{n} -  \bar{n}_A ) \left (
{dN^{\Delta B}_{DB} \over dy}(y) \right )_{\bar{n}/\bar{n}_A} \right ] \quad .  			 \eeq 

\noi Here $\bar{n}_A$ is the average number of participants in each nucleus and
$\bar{n}$ the average number of collisions. $dN_{DP}/dy$ is given by the conventional DP
hadronization me\-cha\-nism and $dN_{DB}/dy$ is the rapidity density given 
by the DB component
--which behaves \cite{7r,9r,10r} as $\exp [- {1 \over 2} |y_{\Delta B} -
y_{Max}|]$.\footnote{This is the behaviour in the original work \protect{\cite{7r}} and is used by
all authors at present energies \cite{10r,11r}. However, a flat behaviour in $y$ at high energies has
been proposed in \cite{8r}, \cite{9r}. This has very important consequences for the net baryon
number at $y^* \sim 0$ at LHC \cite {11r}. A new component consisting on a diquark which contains a
quark from the nucleon sea also has been introduced in \cite{13r}. However, in this case the diquark
hadronizes as a valence diquark producing baryons mainly in the fragmentation regions.} Both
distributions depend, due to energy-conservation, on the average number of collisions per nucleon
$\overline{n/n_A} \approx \bar{n}/\bar{n}_A$. Note that for $\bar{n} = \bar{n}_A$, only the DP
component is present and we recover the usual scaling in the number of participants. Baryon number
conservation implies that both DP and DB rapidity distributions are normalized in such a way that
their integral over $y$ is equal to two. Following \cite{11r} we have 

\beq
\label{2e}
{dN_{DB}^{\Delta B} \over dy} (y) = C_{\bar{n}/\bar{n}_A} \left ( Z_+^{1/2} (1 -
Z_+)^{\gamma_{\bar{n}/\bar{n}_A}} + Z_-^{1/2} (1 - Z_-)^{\gamma_{\bar{n}/\bar{n}_A}} \right ) 
\eeq   
	
\noi where $Z_{\pm} = \exp (\pm y - y_{max})$. The same shape will be used for all baryon species.
Two different ansatzs for the power $\gamma$ were proposed in \cite{11r} --which give for central $PbPb$
collisions $(\bar{n}/\bar{n}_A \sim 4)$ a value of $dN_{DB}^{\Delta B}/dy$ at $y^* \sim 0$ in the range 0.5
to 0.60. In order to reproduce the data of the NA49 collaboration \cite{1r} for $B$-$\bar{B}$ at
$y^* = 0$, we need a value 0.48. Our analysis will concentrate on mid-rapidities and this value
will be used throughout this paper. \par 

Eq. (\ref{1e}) also applies to $pp$ collisions putting $n = n_A = 1$. However, this applies only
at low energies where the two string configuration dominates. At high energies the second term in
(\ref{1e}) is present and increases the net baryon stopping. Unfortunately, the only available data
at ISR do not allow to either confirm or rule out the presence of the DB
component. However, the effect of this component is quite important in $pA$ collisions where the
$pPb$ data of \cite{1r} confirm its presence (see below). In DPM, the extension of eq. (\ref{1e})
to $pA$ collisions is straightforward. A good approximation to the exact formula, valid at $y^* =
0$, is 

\beq
\label{3e}
{dN^{pA \to \Delta B} \over dy}(0) = {\bar{n} \over 2} \ {dN^{pp\to \Delta B} \over dy}(0) + {1
\over 2} \left ( {dN^{\Delta B} \over dy} (0) \right )_{\bar{n}} \quad . \eeq 

\noi Here $\bar{n} = A \ \sigma_{pp}/\sigma_{pA}$ and $(dN^{\Delta B}/dy)_{\bar n}$ is given
by eq. (\ref{1e}) with $n_A = 1$. Eq. (\ref{3e}) is self-evident. On the nucleus side we have
$\bar{n}$ diquarks which hadronizes as in $pp$ collisions (the average number of collisions per
nucleon at present energies is close to one in both cases). On the proton side, we have one
diquark which hadronizes as in $AA$ collisions. (The average number of
collisions of this nucleon in $pPb$ and in central $PbPb$ collisions is practically the same). Eq.
(\ref{3e}) is very useful since it gives the rapidity density of the net baryon in $pPb$, in
terms of the same quantities which appear in $PbPb$ (eq. (\ref{1e})) plus the corresponding one
for $pp$.\par \vskip 5 truemm

\noi {\bf \underbar{B-$\bar{\bf B}$ pair production}.} Following \cite{5r,12r} the rapidity
distributions of antibaryons in $AA$ collisions is given by

\beq
\label{4e}
{dN^{AA \to \bar{B}} \over dy}(y) = \bar{n}_A \left ( {dN^{\bar{B}}_{string} \over dy}(y) \right
)_{\bar{n}/\bar{n}_A} +  (\bar{n} - \bar{n}_A) \left ( {dN^{\bar{B}}_{sea} \over dy}(y) \right
)_{\bar{n}/\bar{n}_A} \quad .
 \eeq

\noi Here $dN_{string}/dy$ denotes the conventional pair production resulting from $d$-$\bar{d}$
production in the string breaking process. At present energies the $q$-$\bar{q}$ strings have too
little energy for baryon pair production~; practically all their production occurs in $qq$-$q$
strings, hence the scaling in the number of participants. This gives a much too small antibaryon
yield. For this reason we have introduced \cite{5r,12r} the second term in (\ref{4e}) which
corresponds to pair production in strings with a $d$ or $\bar{d}$ from the nucleon sea at one
of their ends. The average number of $d$-$\bar{d}$ pairs scales as $\bar{n}$-$\bar{n}_A$. Note
that for $dN_{string}/dy \sim dN_{sea}/dy$, the scaling in the number of participants is
converted into a scaling in the number of collisions. \par

For $pA$ collisions the equivalent of expression (\ref{3e}), valid at $y^* = 0$, is

\beq
\label{5e}
{dN^{pA\to \bar{B}} \over dy}(0) = {\bar{n} \over 2} \ {dN^{pp \to \bar{B}} \over dy}(0) + {1 \over
2} \left ( {dN^{\bar{B}} \over dy} (0)\right )_{\bar{n}} 
\eeq

\noi Here $\bar{n} = A \ \sigma_{pp}/\sigma_{pA}$ and $(dN^{\bar{B}}/dy)_{\bar n}$ is
given by eq. (\ref{4e}) with $n_A = 1$. \par \vskip 5 truemm

\noi {\bf \underbar{Numerical results}.} Let us concentrate first on the
non-conventional contributions (second terms in eqs. (\ref{1e}-\ref{3e}) and (\ref{4e}-\ref{5e}))
--which are nu\-me\-ri\-cal\-ly very important since they contain the factor $\bar{n}$-$\bar{n}_A$.
Actually, $dN_{sea}^{\bar{B}}/dy$ has been computed in \cite{5r}. However, as stated there, there
is a huge uncertainty on its absolute normalization. In view of that, we are going to determine it
from $pA$ data. Since data \cite{2r} for multi-strange hyperons are only available in the
rapidity range $- 0.5 < y^* < 0.5$ we shall restrict our analysis to this region. \par

Turning to the relative yields of the different baryon species, we note that in
$dN_{sea}^{\bar{B}}/dy$ the baryons and antibaryons are produced from three sea quarks. Therefore
the factors controlling the relative yields are~: $4L^3$, $4L^3$, $12L^2S$, $3LS^2$, $3LS^2$
and $S^3$ for $p$, $n$, $\Lambda +\Sigma$, $\Xi^{0}$, $\Xi^-$ and $\Omega$, respectively.
Moreover, we use for both baryons and antibaryons the conventional relation $\Sigma^+ + \Sigma^- =
0.6 \Lambda$, which is obtained after resonance decay. We take $L = 0.435$ and $S = 0.13$. In this
way the strangeness suppression factor has the conventional value  $S/L = 0.3$ and the sum of all
baryon yields add up to $(2L + S)^3 = 1$. However, this fixes only the relative normalization of the
various (anti) baryon species --and there is one free parameter which provides the absolute
normalization which is fixed by the data. The obtained values are given in Table I. \par

The situation is quite different for the new component $dN^{\Delta B}_{DB}/dy$. Here net baryon
conservation fixes the absolute normalization and therefore there is in principle no free
parameter --since the strangeness suppression factors are the same as above. However, it turns out
that proceeding in this way the strange baryon yields in $pA$ are
overestimated --especially $\Xi$'s and $\Omega$'s. A possible explanation of this fact is
that, at present energies, it may happen that the baryon is not formed out of three sea quarks as
in Fig.~1, but, due to lack of phase space, the valence quark at the string end is picked up
instead\footnote{Nevertheless, the factors given above should apply at very high
energy. They give a ratio $\Xi/\Lambda+\Sigma \sim 0.3$ in agreement with Tevatron \cite{14r} and SPS
collider \cite{15r} data.}. In this case, the relative yields of $\Lambda$'s of $\Xi$'s will be
reduced and that of $\Omega$'s will be zero. In any case, we take here a phenomenological attitude.
We determine the strange net baryon yields at $y^* = 0$ from the data. We have thus three free
parameters $\Lambda$-$\bar{\Lambda}$, $\Xi^-$-$\bar{\Xi}^+$ and $\Omega$-$\bar{\Omega}$ --plus a
forth parameter for the absolute normalization of antibaryons as discussed above. We arrive in
this way to the values listed in Table I.  \par

Finally, the conventional components $dN_{DP}/dy$ and $dN^{\bar{B}}_{string}/dy$ have been obtained
in two different ways~: using the formalism of ref. \cite{5r} and using the QGSM \cite{16r,17r} Monte
Carlo without string fusion\footnote{The parameters of this Monte Carlo, which produced too many
strange baryons in $pp$ collisions \cite{17r}, have been tuned up, and its results in $pp$
collisions are now in agreement with the calculations of \cite{5r} and with the experimental data
(see Table 1 in \protect{\cite{5r}}).}. In the former case, the obtained values are given in Table
I. It should be noted that the contribution of these conventional components to the final results is
significantly smaller than the one of the new components, since the latter are proportional to
$\bar{n}$-$\bar{n}_A$. \par

With the values given in Tables I we obtain the results shown in Figs.~3 and 4. The $pA$ data are
well reproduced for all baryon species. In $PbPb$, the $p + \bar{p}$ and most of the $\Lambda +
\bar{\Lambda}$ yield are reproduced. However, the $\Xi^- + \bar{\Xi}^+$ yield is somewhat
underestimated and the $\Omega + \bar{\Omega}$ one is five times smaller than the measured
one\footnote{A qualitatively similar result has been obtained in \protect{\cite{18r}}. This paper,
which was appeared after completion of the present work, is based on a different realization of
baryon stopping and uses a junction-antijunction exchange mechanism for baryon pair production. It
does not include final state interaction.}. \par

In an attempt to solve this problem, we use our expressions (\ref{1e}) and (\ref{4e}) as
initial baryon densities for final state interaction. We consider the following
channels

\beq
\label{6e}
\pi N \to K\Lambda \ , \quad \pi N \to K\Sigma \ , \quad \pi \Lambda \to K\Xi \ , \quad \pi \Sigma
\to K \Xi \ , \quad \pi \Xi \to K \Omega  \quad ,\eeq

\noi plus the corresponding ones for antiparticles\footnote{As a technical point, one should
notice that some of the charge combinations in (\protect{\ref{6e}}) are not possible or have
negligeably small cross-sections. The latter correspond to quark diagrams with three quark lines
(baryon exchange) in the $t$-channel. They have not been included. All reactions corresponding to
quark diagrams with light quark annihilation and $s$-$\bar{s}$ production \protect{\cite{3r}} have
been included with identical cross-section. We have neglected all strangeness exchange reactions
$\pi\Lambda \to KN$, etc. Although, at threshold, the cross-sections are larger than those of the
non-strangeness exchange reactions (\protect{\ref{6e}}), this is no longer the case for the averaged
cross-sections $<\sigma >$ (see \protect{\cite{3r}}). Channels (\protect{\ref{6e}}) are thus
dominant due to the relations $\rho_N > \rho_{\Lambda} > \rho_{\Xi} > \rho_{\Omega}$ between baryon
densities.}. The final state interaction is treated in the same way as in ref. \cite{19r,20r}. The
products of densities in (\ref{6e}) are taken at fixed values of $\vec{b}$ and $\vec{s}$. For the
baryon densities, the dependence on these variables is contained in the geometrical factors
$\bar{n}_A$ and $\bar{n}$ in eqs. (\ref{1e}) and (\ref{4e}). They are computed in the Glauber model
with Saxon-Woods profiles. The pion densities and the initial and final times are the same as in
\cite{19r,20r}. The only difference resides in the fact that, in the case of $J/\psi$ suppression,
there is a single channel and the $J/\psi$ appears both in the initial and final state. In this
case, the differential equations for the gain and loss of particles can be solved analytically and a
simple exponential form is obtained. In the present case, the gain and loss differential equations
for the solution of the coupled channels (\ref{6e}) has to be obtained numerically. We have only one
free parameter~: the cross-section $<\sigma >$ averaged over the momentum distribution of the
incoming particles which is taken to be the same for all reactions (the relative velocity between
the interacting particles is included in $<\sigma >$). We take $<\sigma > = 0.14$~fm. (A
comparable value has been obtained in \cite{3r}).  It is clear that the relative enhancement of the
various baryon and antibaryon yields resulting from the interactions (\ref{6e}) will be the largest
for $\Omega$'s and the smallest for $\Lambda$'s. This is due to the fact that $\Lambda$'s are
produced in some reactions and destroyed in others. On the contrary $\Omega$'s are produced and never
destroyed. The results are shown in Figs.~3 and 4. A reasonable description of all baryon and
antibaryon is obtained. However, the ratio $\Omega + \bar{\Omega}/\Xi^- + \bar{\Xi}^+$ in $PbPb$
collisions is slightly underestimated. \par \vskip 5 truemm

In conclusion, we have introduced a new formulation of the diquark breaking 
mechanism which describes baryon stopping with a diquark breaking probability
equal to one. Incorporated in DPM, together with strings originating from
diquark-antidiquark pairs from the nucleon sea and final state interactions
we have obtained a reasonable description of all baryon and antibaryon yields
at mid-rapidities in $pp$, $pPb$ and $PbPb$ collisions.

\noi {\bf Acknowledgements.} \par \vskip 3 truemm

It is a pleasure to thank A. Kaidalov for many
useful discussions and suggestions. We also thank C. Pajares for comments 
on the idea of this work. One of the authors (A.C.) thanks J. A. Casado for a correspondence
on subjects related to the present work. He also acknowledges partial support from a NATO grant
OUTR.LG 971390. E.G.F. thanks Fundaci\'on Ram\'on Areces 
from Spain and C.A.S Fundaci\'on 
Caixa Galicia  from Spain for financial support. 

\newpage
\section*{Figure Captions}
\vskip 1 truecm
\begin{description}
\item{\bf Fig. 1 :} Diquark breaking component for nucleon-nucleus scattering and two inelastic collisions.
\item{\bf Fig. 2 :} DP component for nucleon-nucleus scattering for two inelastic collisions.
\item{\bf Fig. 3 :} Baryon yields for $pPb$ and $PbPb$ collisions at 160 GeV per nucleon calculated
from eqs. (\ref{1e}-\ref{5e}) using the values in Table 1 before (dotted line) and after (full line)
final state interaction. The dashed lines are the results with final state interaction using
the Monte-Carlo results \protect{\cite{16r}} \protect{\cite{17r}} for the conventional DP and string
breaking components. The data are from the WA 97 collaboration \protect{\cite{2r}}. 
\vskip 5 truemm
\item{\bf Fig. 4 :} Same as in Fig. 3 for the ratios $\bar{B}/B$.
\end{description}	
\vskip 2 truecm
\newpage
\section*{Table}
\vskip 1 truecm
\begin{center}
\begin{tabular}{|l|c|c|c|c|}
\hline
& & & & \\
&$p$ &$\Lambda$ &$\Xi^-$ &$\Omega$ \\
& & & & \\
\hline
& & & & \\
$dN_{DB}^{\Delta B}/dy$ &$1.87 \times 10^{-1}$ &$6.22 \times 10^{-2}$ &$3.26 \times 10^{-3}$
&$2.38 \times 10^{-4}$ \\
$dN_{sea}^{\bar{B}}/dy$ &$4.36 \times 10^{-3}$ &$2.44 \times 10^{-3}$ &$2.92 \times 10^{-4}$ &$2.91
\times 10^{-5}$ \\
$dN_{DP}^{\Delta B}/dy$ &$6.90 \times 10^{-2}$ &$1.40 \times 10^{-2}$ &0 &0 \\
$dN_{string}^{\bar{B}}/dy$ &$8.50 \times 10^{-3}$ &$2.83 \times 10^{-3}$ &$1.65 \times 10^{-4}$
&$5.05 \times 10^{-6}$ \\  $dN_{pp}^{\Delta B}/dy$ &$3.60 \times 10^{-2}$ &$7.00 \times 10^{-3}$
&0 &0 \\ $dN_{pp}^{\bar{B}}/dy$ &$1.70 \times 10^{-2}$ &$5.65 \times 10^{-3}$ &$3.30 \times 10^{-4}$
&$1.01 \times 10^{-5}$ \\
& & & & \\
\hline
\end{tabular}
\end{center}
\vskip 1 truecm
\noi {\bf Table I :} Values of the rapidity densities at $y^* = 0$ in eqs.
(\ref{1e},\ref{3e}-\ref{5e}). The values in the first lower rows are for central $PbPb$ collisions
($\bar{n}_A = 178$, $\bar{n} = 858$) and those in the last two rows for $pp$. Three numbers in the first
row and one in the second one have been adjusted to the data (see main text).

\newpage

\centerline{\bf Figure 1}
\vspace{1cm}

\begin{center}
\hspace{-1.2cm}\epsfig{file=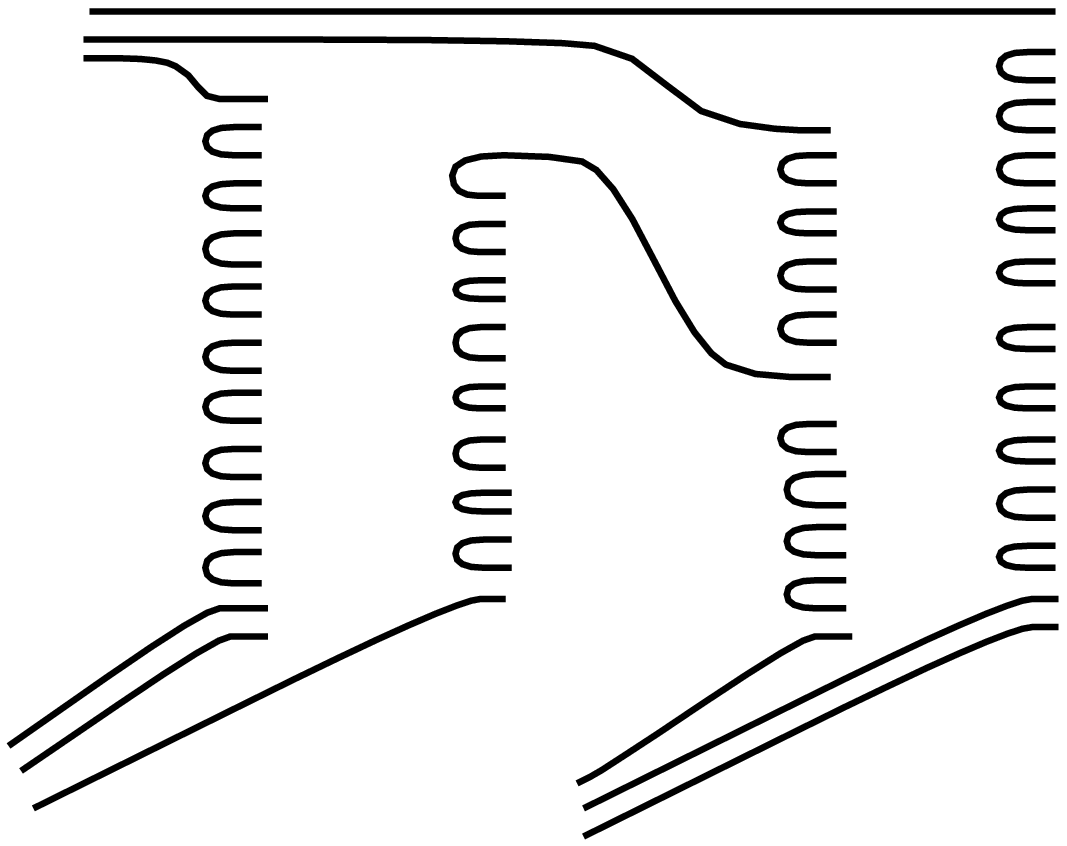,width=10.cm}
\end{center}

\newpage

\centerline{\bf Figure 2}
\vspace{1cm}

\begin{center}
\hspace{-1.2cm}\epsfig{file=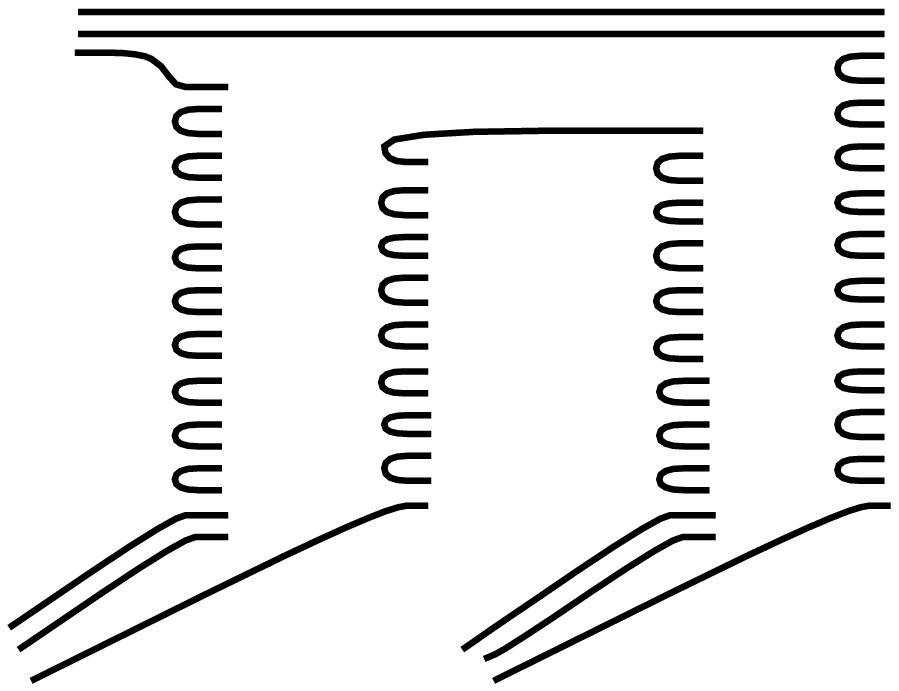,width=10.cm}
\end{center}

\newpage

\centerline{\bf Figure 3}
\vspace{1cm}

\begin{center}
\hspace{-1.2cm}\epsfig{file=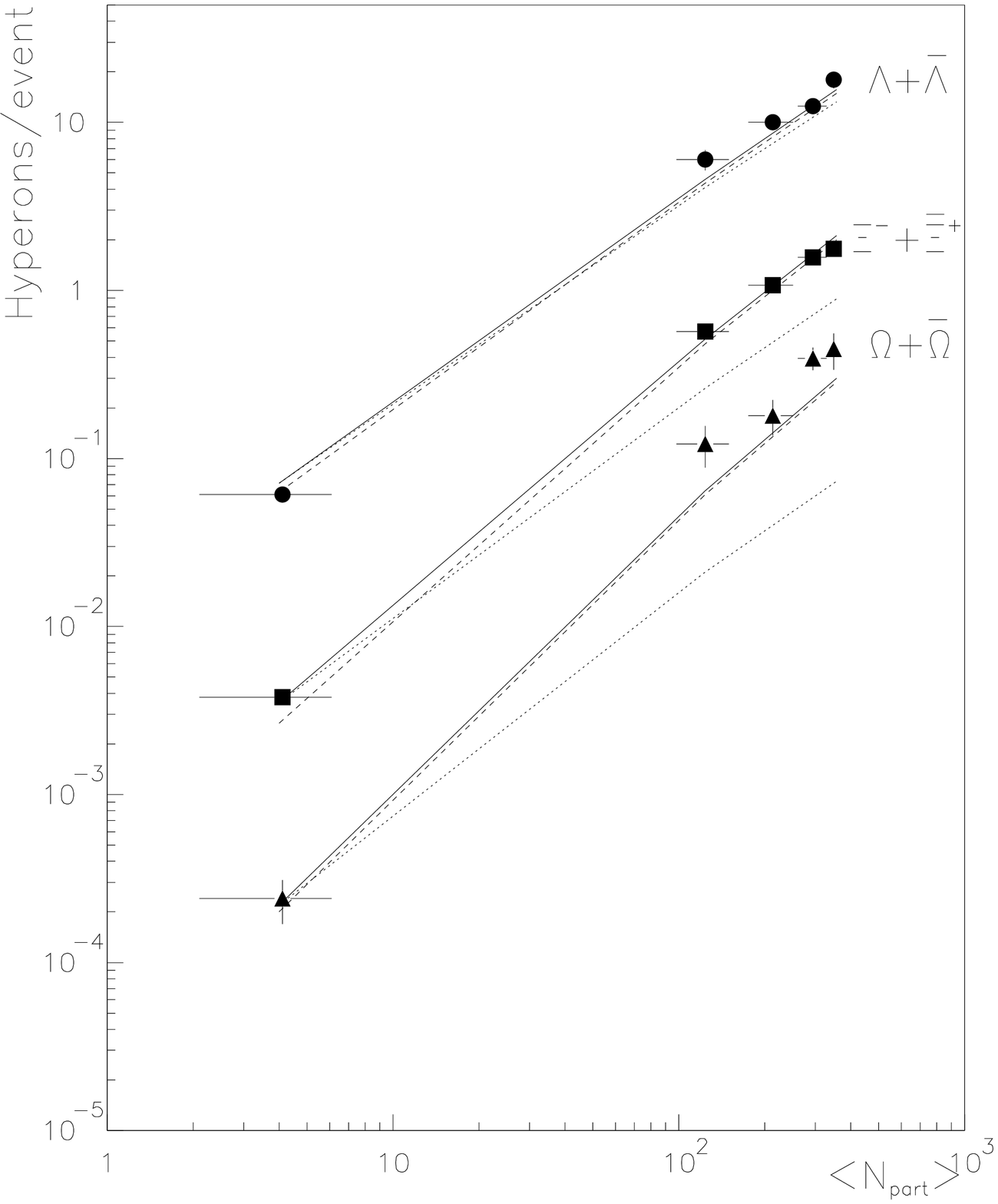,width=13.cm}
\end{center}

\newpage

\centerline{\bf Figure 4}
\vspace{1cm}

\begin{center}
\hspace{-1.2cm}\epsfig{file=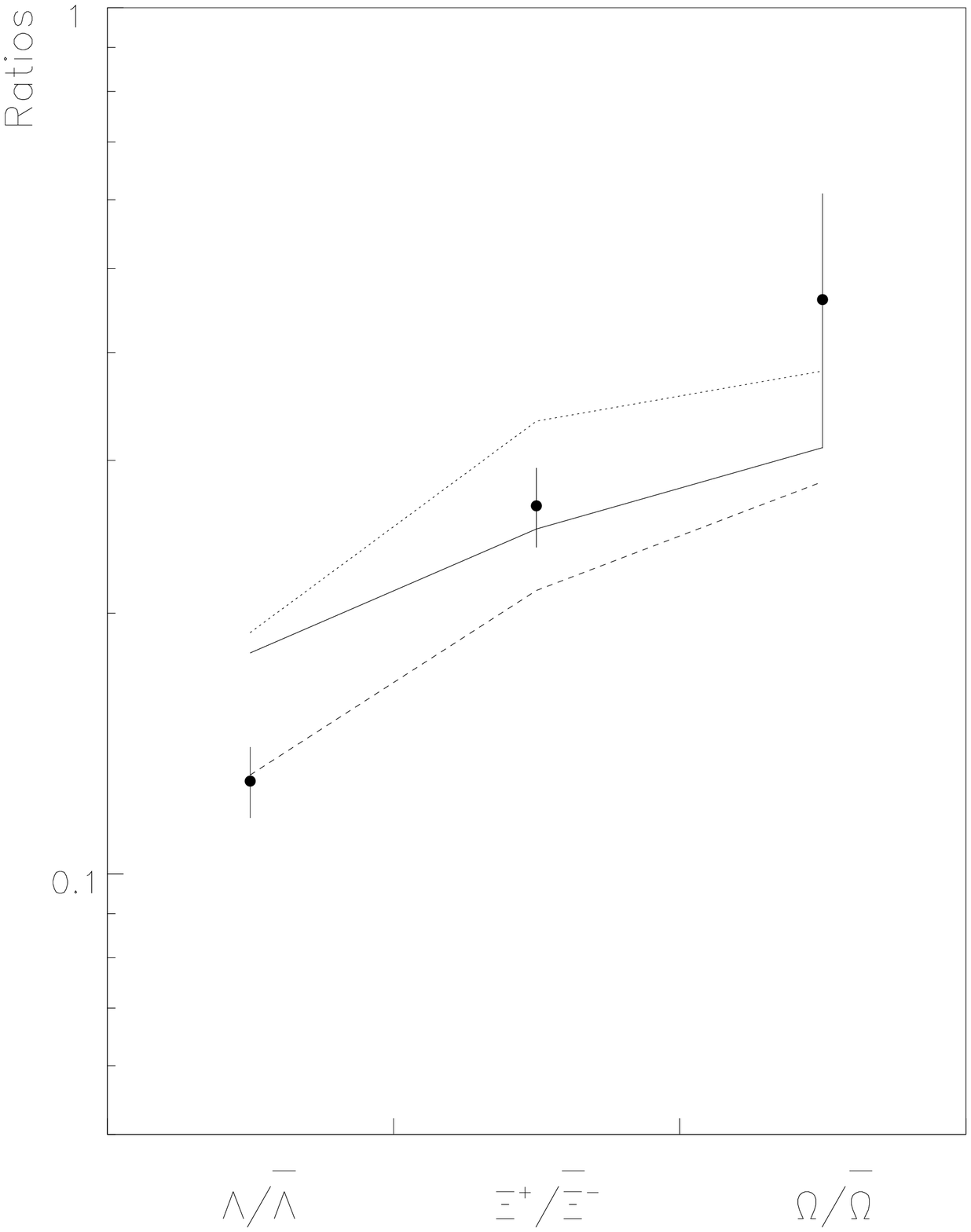,width=13.cm}
\end{center}

\end{document}